\documentclass[a4paper,twoside,openright]{article}
\usepackage{geometry}
\usepackage[latin1]{inputenc}

\usepackage{amsmath}
\usepackage{amssymb}

\usepackage{fancyhdr}
\usepackage[hang,bf]{caption2}

\usepackage{array}
\usepackage{color}
\usepackage{graphicx}

\title{{\b
{Order in extremal trajectories.}
}}

\author{
  Khanh-Dang {Nguyen Thu Lam} \\
  M2 SDSMC--PTSC, Universit\'e Paris VII, \\
  ~\\
Jorge Kurchan \\
CNRS-ESPCI, rue Vauquelin 10, 75231 Paris, France,\\
~\\
and \\
~\\
Dov Levine,\\
Department of Physics, Technion, Haifa 32000, Israel.
}

\begin{document}
\nocite{*}
\maketitle

\vspace{1in}

\begin{abstract}
Given  a chaotic dynamical system  and a time interval in which some quantity takes an unusually large average value,
what can we say of the trajectory that yields this deviation?
As an example, we study the trajectories of the archetypical  chaotic system, the baker's map.  We show that, out of all irregular trajectories, 
a large-deviation requirement selects (isolated) orbits that are periodic or quasiperiodic.
We discuss what the relevance of this calculation may be for dynamical systems and for glasses.
\end{abstract}

\thispagestyle{empty}

\clearpage

\clearpage


\section{Introduction}

In this paper we discuss, with the aid of a simple example, how order arises from chaos 
when we select the solutions that  minimize some quantity.
The motivation for this work is twofold, one related to dynamical systems, and
the other to glasses.

Consider first a chaotic dynamical system, for example a liquid flowing past an obstacle, at large Reynolds numbers.
The wake behind the obstacle is chaotic, and the 
 pressure it creates  may exhibit considerable fluctuations.
We may ask the question: given that during a certain time-interval the average pressure 
has been abnormally large, what can we say about the velocity field during that time?
Is it reasonable to assume that, if the high pressure is to last long, then the fluid phase-space 
trajectory has to have during that time some regularity
that is absent in a typical chaotic trajectory?
What we are asking, then, is a question on the large deviation functions of a chaotic system:  
are the trajectories that sustain extreme situations special, and in a sense simpler?

A second motivation is related to the question of the nature of the  order in an ideal, equilibrium,  glass state
-- if, of course, such a thing exists. An approach to the glass problem~\cite{density_functional} 
is to consider a free-energy density functional in $d$-dimensional
space~\cite{RY}:
\begin{equation}
F[\rho({\bf x})] = \int d^d {\bf x} \; \rho[\ln \rho({\bf x}) -1] - \frac{1}{2} \int  d^d{\bf x} \; 
d^d {\bf x'} \; [\rho({\bf x})-\rho_o] C({\bf x}-{\bf x'})  [\rho({\bf x'})-\rho_o]
\label{free_ene}
\end{equation}
where $C({\bf x}-{\bf x'},\rho_o) $ is the liquid direct correlation function at average density $\rho_o$.
We look for the 'local' free energy minima that  satisfy:
\begin{equation}
\frac{\delta F[\rho({\bf x})]}{\delta {\bf x}} =\ln \rho({\bf x}) - \int   d^d {\bf x'} \;  C({\bf x}-{\bf x'},\rho_o)  [\rho({\bf x'})-\rho_o]=0
\label{eq_mot}
\end{equation}
At low average densities $\rho_o$, the spatially constant 'liquid' solution dominates. As the density increases, a periodic, 'crystalline' solution
appears. What is interesting from the glassy point of view, is that in the high density regime, there appear also many non-periodic 'amorphous'
solutions. Each one of these is supposed to represent a metastable glassy state.
These states are local minima of (\ref{free_ene}) satisfying (\ref{eq_mot}). Beyond the  mean-field  approximation, high free-energy solutions are unstabilized
by nucleation;
if we wish to model the realistic situation we should look for the solutions that are deepest in free energy -- excluding, of course, the crystalline state.
We have then a situation analogous  to the one described above for chaotic systems: of all the solutions satisfying the  "equations of motion"  (\ref{eq_mot}), we have to look 
for the  configurations  that minimize globally the free energy (\ref{free_ene}), and  ask whether these "large deviation" solutions have special regularity properties. 

This analogy between  dynamical systems that are chaotic in time, and glassy systems that are  chaotic in space, was pointed out many years ago by Ruelle~\cite{Ruelle}.
It has also appeared in the theory of charge-density waves  \cite{Aubry,CF}, in particular in the Frenkel-Kontorova model.
It turns out that the local energy minima of the model are given by the trajectories of the "standard map", which has both regular and chaotic orbits.
However, when one restricts to the lowest  energy minima, the trajectories selected are quasiperiodic.

In a recent paper it was argued~\cite{KL} that ideal glasses should exhibit a form of 
 geometrical order that is more general than periodic and quasiperiodic,  characterized by the fact that local motifs  of all
 sizes are repeated often. For a system that is not completely ordered, this assumption leads to the definition of a correlation length, defined
 as the largest size of patterns that repeat more frequently than they would in a random case.  
  The claim is then that this 
correlation length should diverge in an ideal glass. But frequent repetitions of
patterns would not occur for a generic solution of a spatially chaotic glass. If order in the solution is to happen at all 
it is   through a mechanism as described above:   
 choosing configurations that globally minimize   energy would then force the system to favor some patterns
that would consequently be repeated often.

In the present work we treat a  simple example.
The system studied  has the advantage that it is purely chaotic, we know that there are no islands of regular dynamics in phase-space.
If the large-deviation bias selects orbits which have some regularity, it does so  in the most unfavorable situation in which these orbits are isolated and unstable.
The organization of this paper is as follows: In
section~\ref{section:model}, we present the baker's map and the functional
used in our model. In section~\ref{section:map}, the  trajectories that extremize 
 the functional are found and their order properties are
discussed.

\section{\label{section:model}The model}

In this section we briefly review the baker's map and some of its
properties. We then define the functional used in our model and whose
minimal trajectories will be studied later.

\subsection{The baker's map}

\begin{figure}[h]
\begin{center}
\includegraphics[width=10cm]{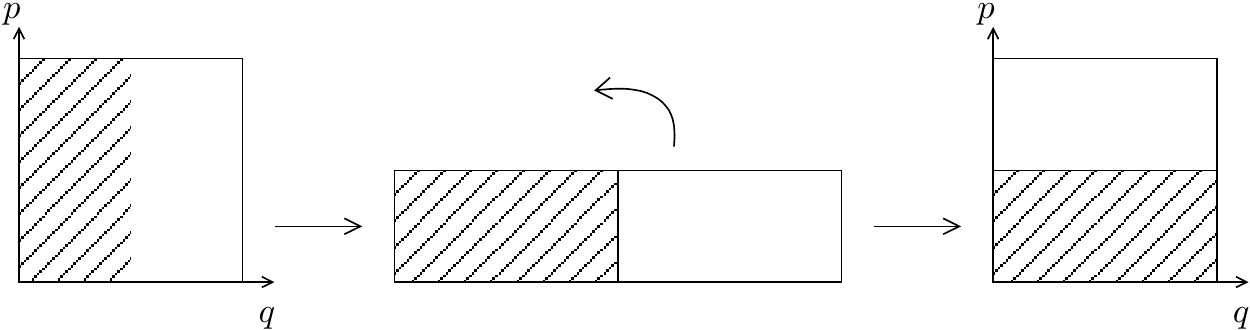}
\end{center}
\caption{\label{fig:bakery}
One iteration of the baker's map. }
\end{figure}

The baker's map is a simple two-dimensional chaotic map acting on the unit
square.  The dynamical system with discretized time $t$, associated with the
baker's map, is defined for $q_t \in [0,1]$ and $p_t \in [0,1]$  by the
area-preserving equations
\begin{eqnarray}
\label{def:bakerx}
q_{t+1} &=& 2 q_t - \lfloor 2 q_t \rfloor , \\
\label{def:bakery}
p_{t+1} &=& \frac12 \left( p_t + \lfloor 2 q_t \rfloor \right) ,
\end{eqnarray}
where $\lfloor 2 q_t\rfloor$ is the integer part of $2q_t$.
As shown by figure~\ref{fig:bakery}, the phase square is
stretched in the $q$ direction and squeezed it in the $p$ direction; 
the baker then cuts the right part and puts it on top of the left one.

The baker's map can also be understood as a shift operator.  Let us consider
the binary representations $q_t = 0.abc\dots$ and $p_t = 0.uvw\dots$ where
$a$, $b$, $c$, $u$, $v$, $w$, \dots are either
$0$ or $1$.  If the position $(q_t,p_t)$ of the system at time $t$ is
written as the string ``$\dots wvu \cdot abc\dots$'', the state of system at
time $t+1$ is obtained by shifting the central dot to the right by one
position, that is, the string becomes ``$\dots wvua \cdot bc\dots$'' and we
have $q_t = 0.bcd\dots$ and $p_t = 0.auv\dots$.

The baker's map is deterministic: if the system is started at initial
conditions  $(q_0,p_0)$, equations~\eqref{def:bakerx} and
\eqref{def:bakery} determine its position at all future (and past) times.
However, the system is chaotic in the sense that  a small uncertainty in
the initial conditions grows exponentially in time (the shift operator
doubles it at each iteration); when the uncertainty reaches the size of the
phase square (here, this size is unity), the position of the system becomes
unpredictable.

\begin{figure}[h]
\begin{center}
\includegraphics[width=10cm]{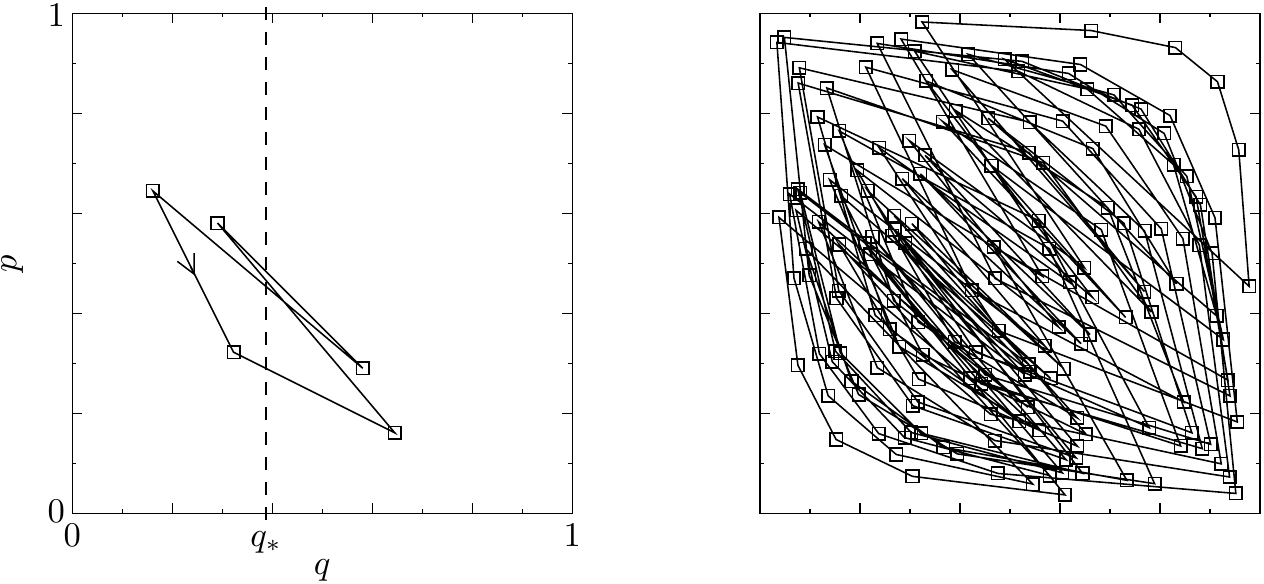}
\end{center}
\caption{\label{fig:orbit5}
Representation of the periodic orbit $\{n_4,\dots,n_0\} = \{0,0,1,0,1\}$ in
the phase square (left). The arrow shows the direction of the trajectory.
An irregular orbit that never is an extremal trajectory, for any $q_*$ (right).
}
\end{figure}

Although almost all trajectories are aperiodic, there are  many 
initial conditions leading to periodic orbits. 
Considering the shift operator interpretation and remembering that the
binary representation of a rational number has a periodic pattern, one sees
that the periodic orbits are related to rational numbers.
In figure~\ref{fig:orbit5} (left), the rational number $5/31$ ( whose binary representation
is the periodic string $0.\,00101\,00101\,00101\dots$) was chosen as
the initial position $q_0$ (the top-left point). The ordinate $p_0$ is
such that its binary representation is the mirror of $q_0$'s binary
representation. After five units of time, the dot shifts five times to the
right and one has $q_5 = q_0$; thus, the orbit is periodic with period 5. A
periodic orbit can be described by the binary digits of the periodic
pattern in the binary representation.  In our example, the orbit is
$\{n_4,\dots,n_0\} = \{0,0,1,0,1\}$. Obviously, a cyclic permutation of
these five digits doesn't change the orbit.

Strictly speaking, the binary representation of a rational is not always
periodic, but the non-periodic part is located near the dot and has a finite
length. For example, the number $q_0=5/31+1/4$ is written as
$0.\,011\,01001\,01001\dots$ After many iterations of the map, the system
forgets the $1/4$ term in $q_0$ (its second most significant bit), because
it is shifted to less and less significant places in the binary
representation of $p_t$; the initial $p_0$ is forgotten as well.
It is clear that as long as the initial position $q_0$ is rational, the
trajectory falls into a periodic orbit after some time.
If a $q_0$ is chosen randomly and uniformly in the unit interval, it is
irrational. Indeed, although rational numbers are dense in the unit
interval, they form a set with zero measure. As a consequence, typical
trajectories are not periodic. Because the
baker's map is ergodic, such a typical trajectory fills up uniformly the
phase square.
Furthermore, a small perturbation of an initial condition associated with a periodic orbit yields a chaotic orbit.

The baker's map has two unstable fixed points which correspond to periodic orbits
with period one.  These two orbits are $\{0\}$ and $\{1\}$, respectively the
points located at the origin $(0,0)$ and the opposite corner $(1,1)$.

\subsection{Large deviations}

Let us  introduce in our model an observable which is a functional of the
trajectories.  Our choice is arbitrary, we are motivated here by simplicity.
We consider a functional
parametrized by a number $q_*$ in $[0,1]$:
\begin{equation}\label{def:obs}
h[\{q_t\}, q_*] = \lim_{T\to\infty} \;
	\frac{1}{T} \sum_{t=0}^{T-1}
		(q_t - q_*)^2 .
\end{equation}
The observable $h$ can be interpreted geometrically as the average squared
distance over time between the trajectory and a reference vertical line
located at $q_*$, as schematically represented in figure~\ref{fig:orbit5}.
For each value of the parameter $q_*$, there is one particular trajectory
that minimizes the average distance $h$.  
Note that $h$ is invariant by the
transformation $(q_t,q_*)\to(1-q_t,1-q_*)$ so if $\{q_t\}$ is an extremal
trajectory for $q_*\le\tfrac12$, $\{1-q_t\}$ is an extremal trajectory for
$1-q_*\ge\tfrac12$. Hence, we can restrict the study to $0\le q_*\le\tfrac12$.
Although the baker's map is bidimensional, the problem is actually
one-dimensional only, because the dynamics on the $q$ axis are not coupled to
the one on the $p$ axis, as shown by the equation~\eqref{def:bakerx}. The
two-dimensional generalization of the observable~\eqref{def:obs} would be
the average squared distance between the trajectory and a reference {\em point}
located at $(q_*,p_*)$:
\begin{equation}\label{def:obs2}
\lim_{T\to\infty} \;
	\frac{1}{T} \sum_{t=0}^{T-1} \left[
		(q_t - q_*)^2
		+ (p_t - p_*)^2 \right] .
\end{equation}
The extremal trajectories for $(q_*,p_*)$ of the
observable~\eqref{def:obs2} are easily deduced from the extremal
trajectories of the one-dimensional observable~\eqref{def:obs} for the
parameter $\tfrac12(q_*+p_*)$. It will indeed be clear later that the
extremal trajectories are left invariant by the transformation that swaps
the $q$ and $p$ axis; this transformation is equivalent to reversing the
time.  The orbit in figure~\ref{fig:orbit5} is for example the minimal
trajectory for the chosen value of $q_*$ (and actually for any $q_*$ in the
whole range $[0.385, 0.411]$) and has the $(p \leftrightarrow q)$ symmetry, as do the
orbits of figure~\ref{fig:orbits}.

\begin{figure}[h]
\begin{center}
\includegraphics{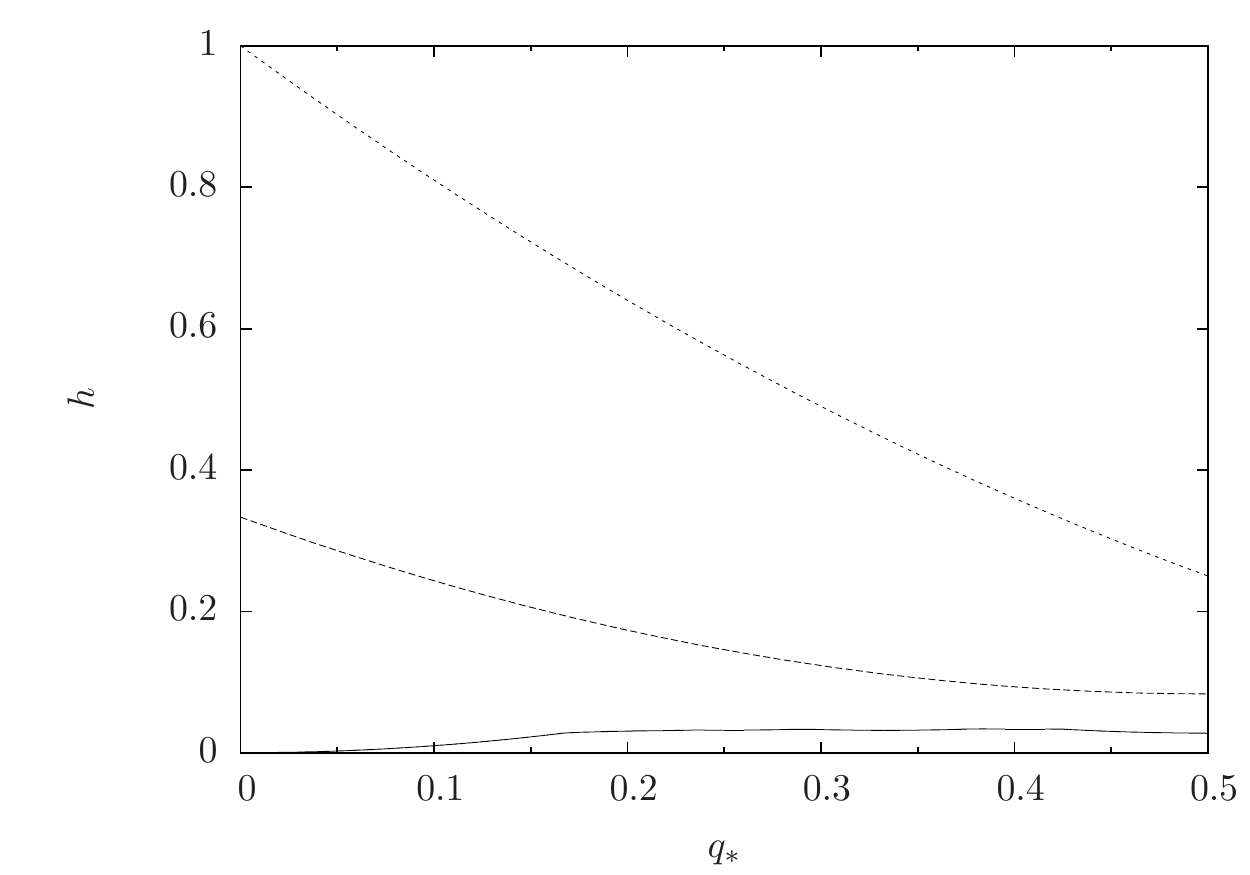}
\end{center}
\caption{\label{fig:spectra}
Value of the observable~\eqref{def:obs} for the minimal trajectory (lowest
curve), the maximal trajectory (highest curve) and a typical
trajectory (middle curve).
}
\end{figure}

The worse state in the sense of minimization of $h$ is one of the two fixed
points. If $q_* \le\tfrac12$, this maximal trajectory is the fixed point
$(1,1)$ and we have $h=(1-q_*)^2$.
 On the other hand, the distribution of the $q_t$ for a chaotic
trajectory is uniform in the unit interval and the mean distance is
calculated by the continuous sum
\begin{equation}
h_\text{chaotic} (q_*)
	= \int_0^1 (q-q_*)^2 \text{d}q
	= \frac13 - q_* (1-q_*).
\end{equation}
Figure~\ref{fig:spectra} shows the energies of the minimal orbit (which we shall determine below), the
maximal orbit and a typical (\textit{i.e.} chaotic) orbit. It is worth
noting that chaotic orbits never minimize or maximize the functional.
In the next sections, the minimal trajectories of $h$ are made explicit and  found
to be periodic for most values of $q_*$ and quasi-periodic for some (zero measure) set of values of $q_*$.

\section{Large-deviation function as a free energy of a  1D lattice gas model%
\label{section:map}%
}

\subsection{Mapping to the particle  model}

In this section, we will show that the functional~\eqref{def:obs} can be
interpreted as the Hamiltonian of  particles  on 
a one-dimensional lattice with single occupancy and a translationally
invariant interaction.  That is,

$$
h =
	\frac{1}{M} \sum_{i,j=1}^M  U_{|i-j|} n_i n_j
	- \mu \rho
	+ h_0 
$$
where $\rho$ is the particle density, and $\mu$ acts as a chemical potential.
The interaction will be shown to decay exponentially with distance.

As stated earlier, a point on a periodic orbit (of period $N$) is represented by the binary sequence
$0.n_{N-1}...n_1n_0n_{N-1}... \equiv \{n_{N-1},\dots,n_1,n_0\}$.  Note that this number may be written as
\begin{equation*}
\{n_{N-1},\dots, n_0\} = \frac{1}{2^{N}-1}
\left( 2^{N-1}n_{N-1} + \dots + 2^1 n_1 + n_0 \right)
.
\end{equation*}

\pagebreak

\pagebreak

We define $\sigma$ as the cyclic left shift operator
(or left cyclic permutation) to be  applied on the digits $n_i$ of the period,
that is
\begin{equation*}
\sigma \{n_{N-1},\dots,n_1,n_0\} = \{n_{N-2},\dots,n_0,n_{N-1}\}
.
\end{equation*}
The $i^{th}$ iteration of $\sigma$ may be written as
\begin{equation*}
\sigma^i \{n_{N-1},\dots,n_0\}
	= \frac{1}{2^N-1} \sum_{j=0}^{N-1} 2^{[j+i]} n_j
,
\end{equation*}
Here and in what follows  $[ \bullet]$ denotes the exponents and indices modulo $N$. 

Taking into account the periodicity and expanding, the observable~\eqref{def:obs} now reads
\begin{equation}\label{eq:obsper}
h =	u - \mu \rho + h_0 ,
\end{equation}
where
\begin{equation}
\rho = \frac1N \sum_{j=0}^{N-1} n_j , \qquad
\mu = 2 q_*  , \qquad
h_0 = q_*^2 .
\end{equation}
and with $u$ given by
\begin{eqnarray}\label{eq:u}
u  &=&
	\frac{1}{N} \sum
		\left( \sigma^i \{n_{N-1},\dots, n_0\} \right)^2= \frac{1}{N (2N-1)^2} \sum_i \sum_{jk} 2^{[j+i]} 2^{[k+i]} \; n_j n_k \nonumber \\
		&=& \frac{1}{N} \sum_{jr} \left( \frac{1}{(2N-1)^2}\sum_i 2^{[j+i]} 2^{[j+r+i]} \right) n_j n_{[j+r]} 
\end{eqnarray}
In the above,  all the indices are summed from $0$ to $N-1$.
Shifting the exponents on the right hand side as $i \rightarrow [i-j]$ we get:
\begin{equation}\label{eq:u1}
u  =\frac{1}{N} \sum_{jr}  U_r^{(N)}  \; n_j n_{[j+r]}   \;\;\;\;  \;\;\;\;  with  \;\;\;\;   \;\;\;\; U_r^{(N)}  = \frac{1}{(2N-1)^2}\sum_i 2^{[i]} 2^{[r+i]}  
\end{equation}
In the Appendix we show that   
\begin{equation}\label{eq:Jr}
U_r^{(N)}= \frac13 \, \frac{2^{N-r} + 2^r }{2^N - 1} .
\end{equation}

\subsection{The long-orbit limit}

\begin{figure}[h]
\begin{center}
 \includegraphics[width=8cm]{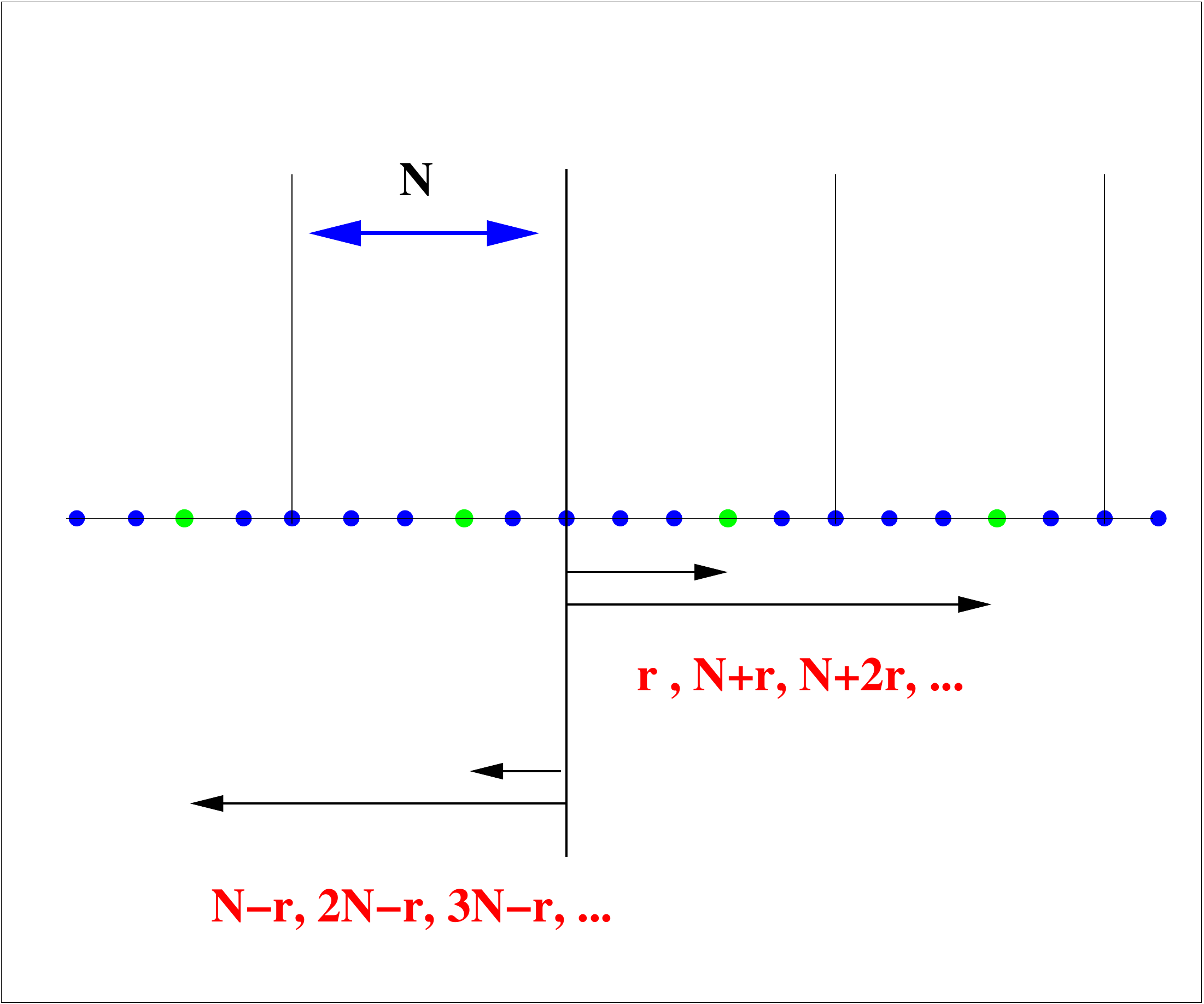}
 \end{center}
\caption{\label{chain}
Counting the energy of a periodic configuration as the sum of two geometric series,
corresponding to the set of periods to the right, and to the left, respectively.
}
\end{figure}

Consider now a very long chain of size $M$, particles with occupation numbers $n_i$, and an interaction given by 
\begin{equation}\label{eq:h1}
h =
	\frac{1}{M} \sum_{i,j=1}^M  U_{|i-j|} n_i n_j
	-  \frac{\mu}{M} \sum_{i=1}^M  n_i 
	+ h_0 =	\frac{1}{3M} \sum_{i,j=1}^M  2^{-|i-j|} \; n_i n_j
	- 2 q_*  \rho
	+ (q_*)^2
\end{equation}
that is,  $U_r= \frac13 2^{-|r|}$.  Let us compute the energy of a  periodic 
configuration of period $N<<M$. We can re-express the unrestricted sum (\ref{eq:h1}) as a sum over the first
period as in Figure \ref{chain}, adding  first each  position $r$ within the first period with its repetitions  in all other periods. We may express  these contributions
as two geometric series, corresponding to the particles to the right and to the left, respectively:
\begin{eqnarray}
 & & \frac{1}{3} \left\{ \left[ 2^{-r} + 2^{-(N+r)} + 2^{-(2N+r)} + ... \right] + \left[ 2^{-(N-r)} + 2^{-(2N-r)} + 2^{-(3N-r)} + ... \right]  \right\} \nonumber \\
&=& \frac{1}{3} \frac{1}{(1-2^{-N})} \left[ 2^{-r} + 2^{-N+r} \right] = \frac{1}{3} \frac{2^{N-r } + 2 ^r}{2^N-1} =U_r^{(N)}
\end{eqnarray}
Thus,  (\ref{eq:u1}) is recovered. This means that we may consider, without loss of generality, a very long periodic chain of size $M$, and the energy of periodic configurations
of periods $N<<M$ will be correctly computed by the interaction (\ref{eq:h1}). This, in turn means that our large functional (\ref{def:obs}) can be computed for large times on the
basis of  the particle model (\ref{eq:h1}).

\subsection{Large deviation function}

Given a time $T$, we  express the probability of observing a value  of the quantity  defined in  (\ref{def:obs})
as:
\begin{equation}
P(h[\{q_t,q_*]=h) = e^{-T \xi(h)} \;\;\;\;\;\;\; ;
\end{equation}
this defines $\xi(h)$.
Alternatively, we may express the same information in the large deviation function $f(\lambda)$ :
\begin{equation}
e^{-Tf(\lambda)} =  \int d\lambda \; P(h[\{q_t,q_*]=h) \; e^{-\lambda T \; h}
\label{ldf}
\end{equation}
For long times $T$, we have that $\xi$ and $f$ are related by a Legendre transform.
In fact, we recognize $f(\lambda)$  as the free energy density of the particle model at inverse temperature $\lambda$.
In particular, the ground state (corresponding to infinite $\lambda$) is given by the trajectory minimizing $h$.

\subsection{The ground state of the lattice particle model}

Let us start by considering the minimum of the energy (\ref{eq:h1}) :
\begin{equation}\label{eq:h2}
\min\{h\} = \min \left\{
	\frac{1}{M} \sum_{jr}  U_{r} n_j n_{[j+r]}
	- 2 q_*  \rho
	+  (q_*)^2 \right\} =  \min_{\rho} \left\{U^{min} (\rho)-2 q_*  \rho
	+  (q_*)^2 \right\}
	\end{equation}
where $U^{min}(\rho)$ is the smallest energy for a fixed  configuration  density.
Although it is tempting to  determine the value of $\rho$ of the extremal configuration by
\begin{equation}
\frac{dU^{min}(\rho)}{d\rho}= 2 q_*
\label{kkk}
\end{equation}
 this must be treated with care, because as we shall see, the derivative is discontinuous. Note again that $q_*$ plays the role of a chemical potential.

Our job is now to determine $U^{min} (\rho)$.
This problem 
 was solved years ago~\cite{hubbard1978,pokrovsky} for  long-range ($U_r>0 \;\; \forall r$) interactions with
$U_r$  convex downwards, as is the case here.

\begin{figure}[htb]
\begin{center}
\includegraphics[width=10cm]{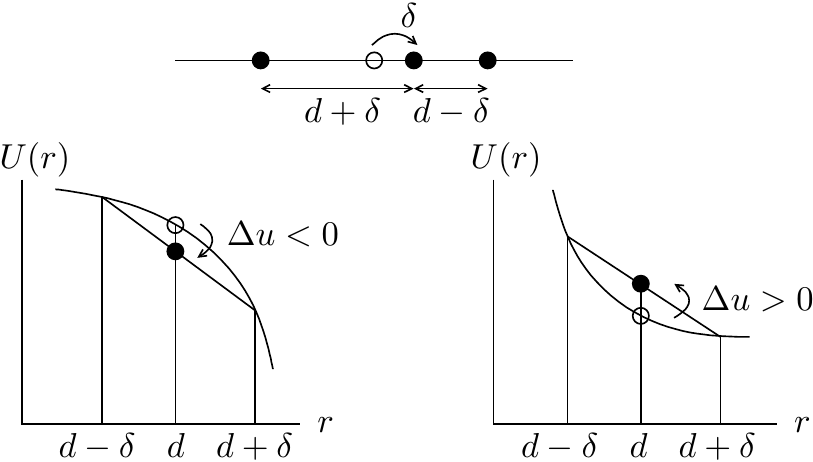}
\end{center}
\caption{\label{fig:convexity}
Change $\Delta u$ of the energy of a cristal configuration after the
displacement of one particle for a concave interaction (left) and a convex
interaction (right).
}
\end{figure}

For the moment, let us forget the lattice.  If the particles can take
 any position on a one-dimensional line with periodic boundary conditions, the ground state is
simply a crystal, that is, each particle is separated from its neighbour by
the same distance $d=1/\rho$.  The condition of convexity of the
interactions can be understood as a stability condition.  Indeed, if one
considers a first-neighbor concave interaction as shown in
Figure~\ref{fig:convexity} (left), a displacement of one particle in the
crystal would decrease the energy of the system: the system is not stable.
On the contrary, if the first-neighbor interaction is convex as shown on the
figure~\ref{fig:convexity} (right), a displacement of one particle would
increase the energy of the system.

\begin{figure}
\begin{center}
\includegraphics[width=4cm]{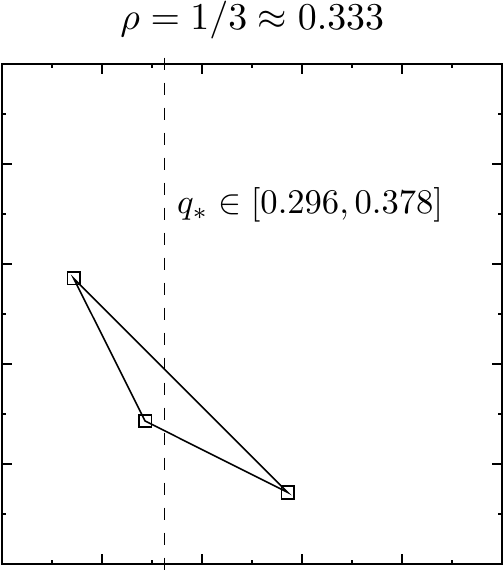}\hspace{20mm}
\includegraphics[width=4cm]{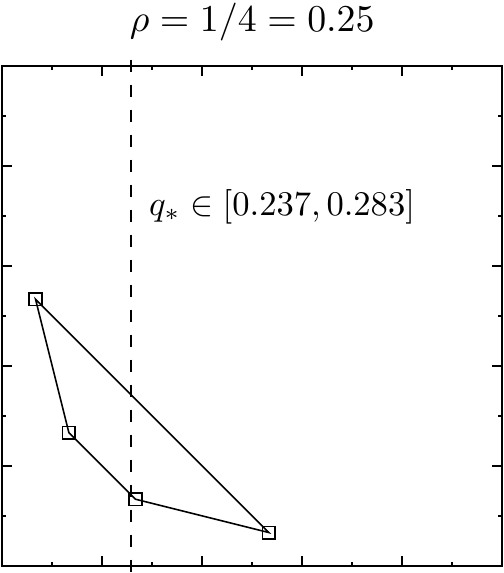} \\[2em]

\includegraphics[width=4cm]{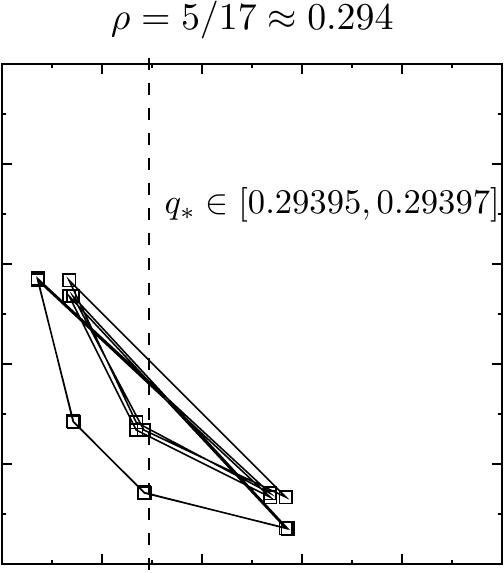}\hspace{20mm}
\includegraphics[width=4cm]{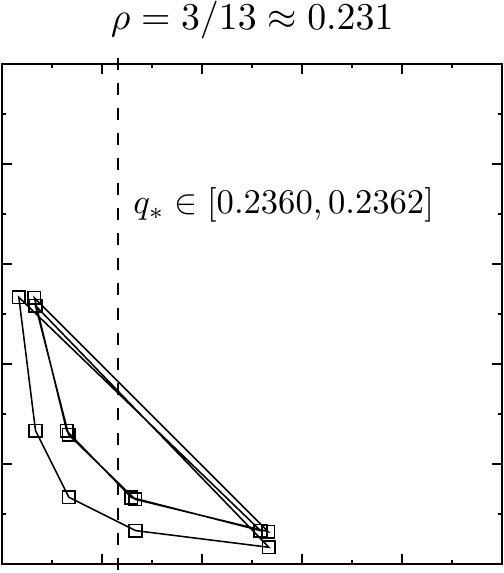} \\[2em]

\includegraphics[width=4cm]{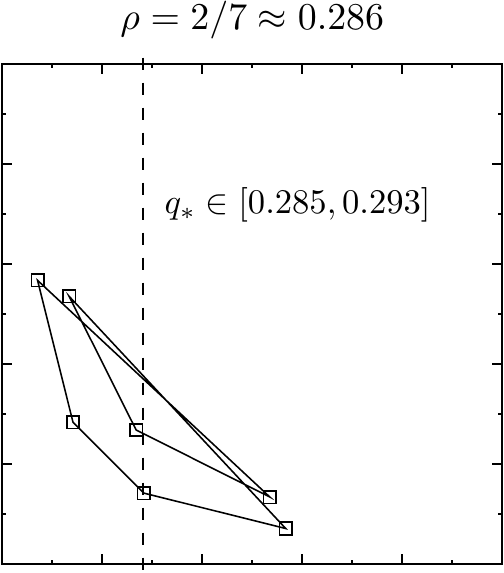}\hspace{20mm}
\includegraphics[width=4cm]{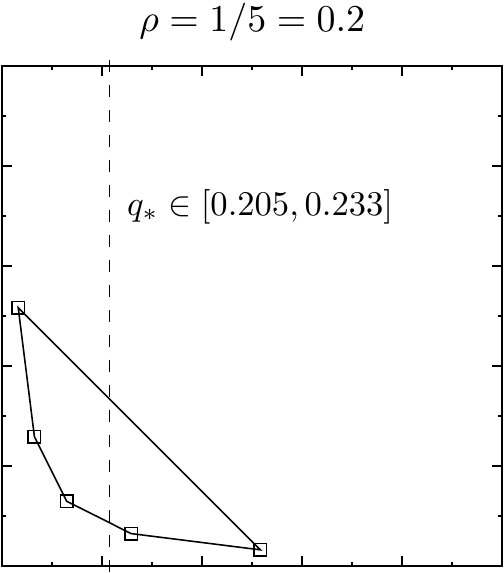}

\end{center}
\caption{\label{fig:orbits}
Examples of minimal orbits for different values of $q_*$. The density of $1$'s 
$\rho$ of the orbits are indicated.
}
\end{figure}

For particles constrained to lie on a lattice, if $\rho$
is such that the mean distance $1/\rho$ is an integer, then the ground state
is again periodic with particles separated by this same distance $1/\rho$.
In the baker's map representation, the trajectory corresponding to such
a crystal is very regular. Figure~\ref{fig:orbits} shows the
$\rho=\tfrac13$, $\rho=\tfrac14$ and $\rho=\tfrac15$ minimal trajectories.
As $1/\rho$ becomes larger, the trajectory spends more and more time near
the fixed point $(0,0)$.  

\begin{figure}
\begin{center}
\includegraphics{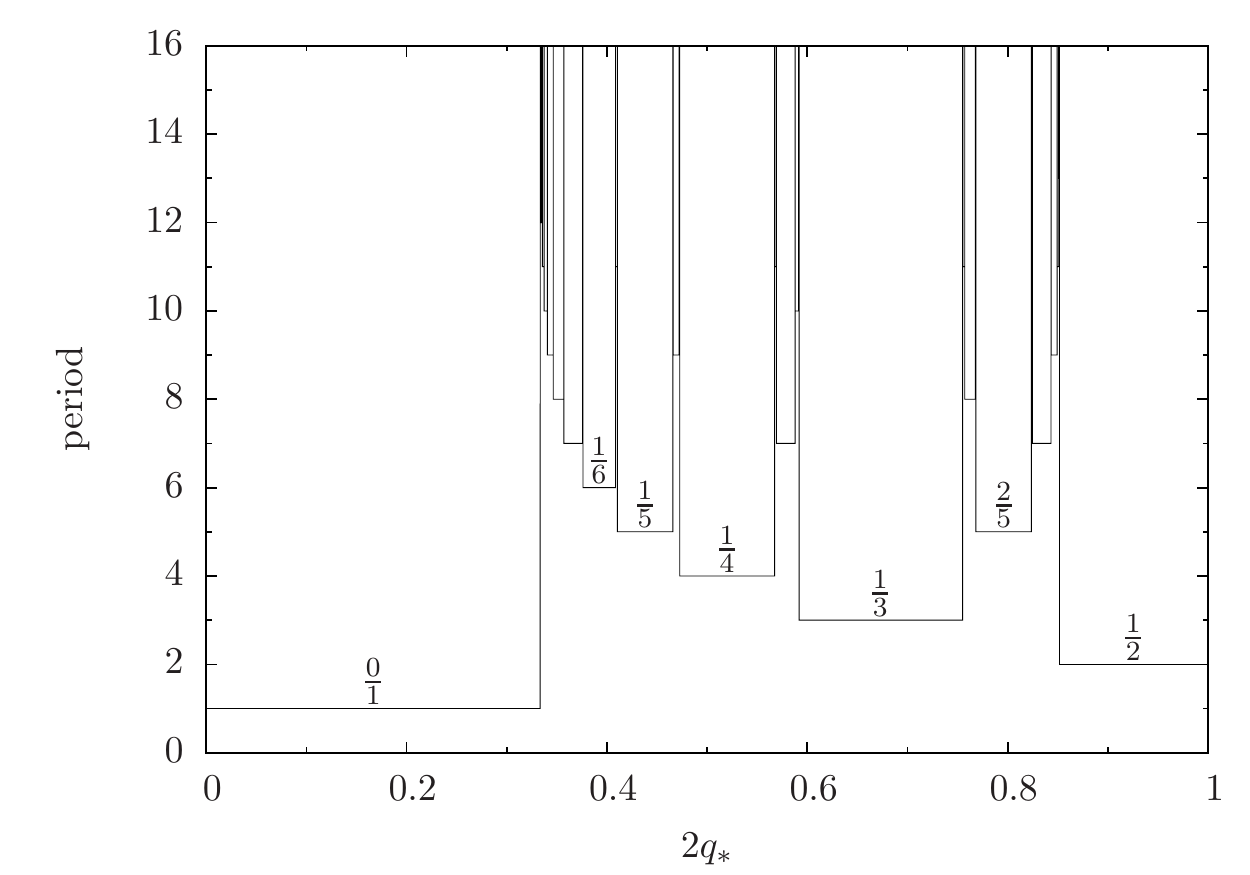}
\end{center}
\caption{\label{fig:period}
The period of the  the minimal trajectory as a function of  $q_*$. The fractions 
specify the density $\rho$ of particles in the corresponding configuration.  Only
the largest steps are indicated.
}
\end{figure}

If $1/\rho$ is not an integer, then it lies between the two successive integers
$d = \lfloor 1/\rho \rfloor$ and $d+1$, and the distance between any two
consecutive particles will be either $d$ or $d+1$.  There is a competition
between the solution with density $1/d$ and the one with density $1/(d+1)$.
This is seen in the baker's map, where Figure~\ref{fig:orbits} shows that the trajectory for
$\rho=5/17$ spends some time in the
$\rho=1/4$  (period $4$) solution and some time in the $\rho=1/3$ (period $3$) solution. %
The competition between the two solutions $1/d$ and $1/(d+1)$ is also seen
in Figure~\ref{fig:period}. In this figure, the $1/d$ solutions are seen to
dominate for some large range of values of $q_*$.  When varying the
parameter $q_*$ continuously in order to go from  the  $1/d$ solution to
the $1/(d+1)$ solution, the period  takes many intermediate values "interpolating" between periods $3$ and $4$.

\begin{figure}
\begin{center}
\includegraphics[width=10cm]{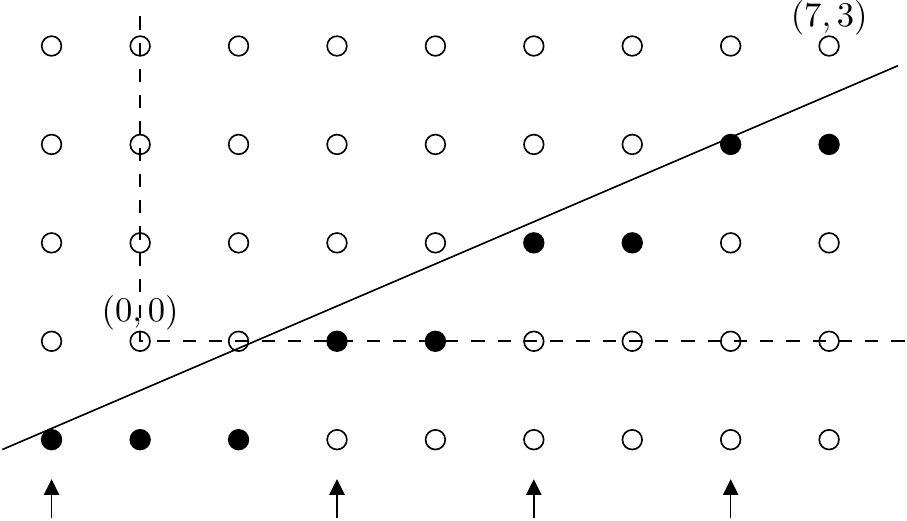}
\end{center}
\caption{\label{fig:bresenham}
Determination of the ground state of the lattice Hamiltonian for $\rho=3/7$. The arrows
indicate the position of the particles.
}
\end{figure}

Let us now  turn to  the full solution, which was 
 originally  given  in terms of continued fractions~\cite{hubbard1978,pokrovsky} and admits a graphical construction related to that employed in quasicrystals\cite{debruijn,steinhardt} which we will describe.
Consider the plane and  the lattice of  points with integer coordinates, as shown in
figure~\ref{fig:bresenham}, and the  line with slope $\rho$ starting
at the origin $(0,0)$.   The algorithm to solve the problem is quite simple: for each integer
abscissa $k$, compute the ideal ordinate $k \rho$ and round it to the
closest smaller integer $y_k$, then, mark the point   $(k,y_k)$.
The ground state corresponding to density $\rho$ is simply given by the
sequence of binary digits $n_k = y_{k}-y_{k-1}$, that is,
there is a particle in site $k$ in places where the black pixels  have a step upwards.
In fact, the line can start at any arbitrary point; this leads to different sequences with the same
energy.  These sequences have the form 
\begin{equation}\label{eq:groundstate}
  n_k = \lfloor k\rho +\phi \rfloor
	- \lfloor (k-1)\rho +\phi \rfloor \;\;\;\;,
\end{equation}
where $\phi$ is a phase factor related to the $y$-intercept of the line.
Figure~\ref{fig:bresenham} shows the determination of the ground state for
$\rho=3/7$. One can read directly the ground state by looking at the
ordinates of the black pixels. 

\begin{figure}
\begin{center}
\includegraphics{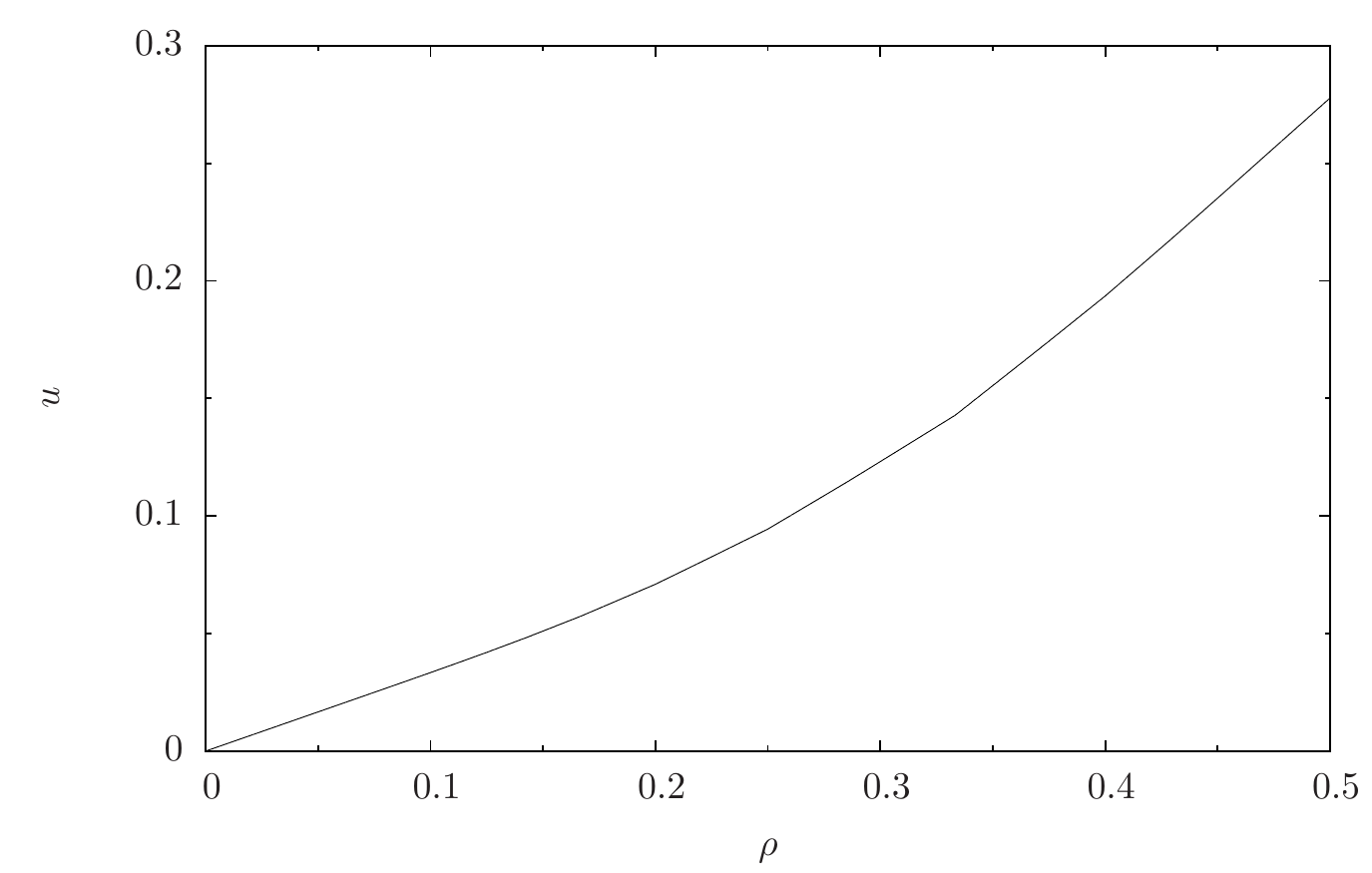} \\
\includegraphics{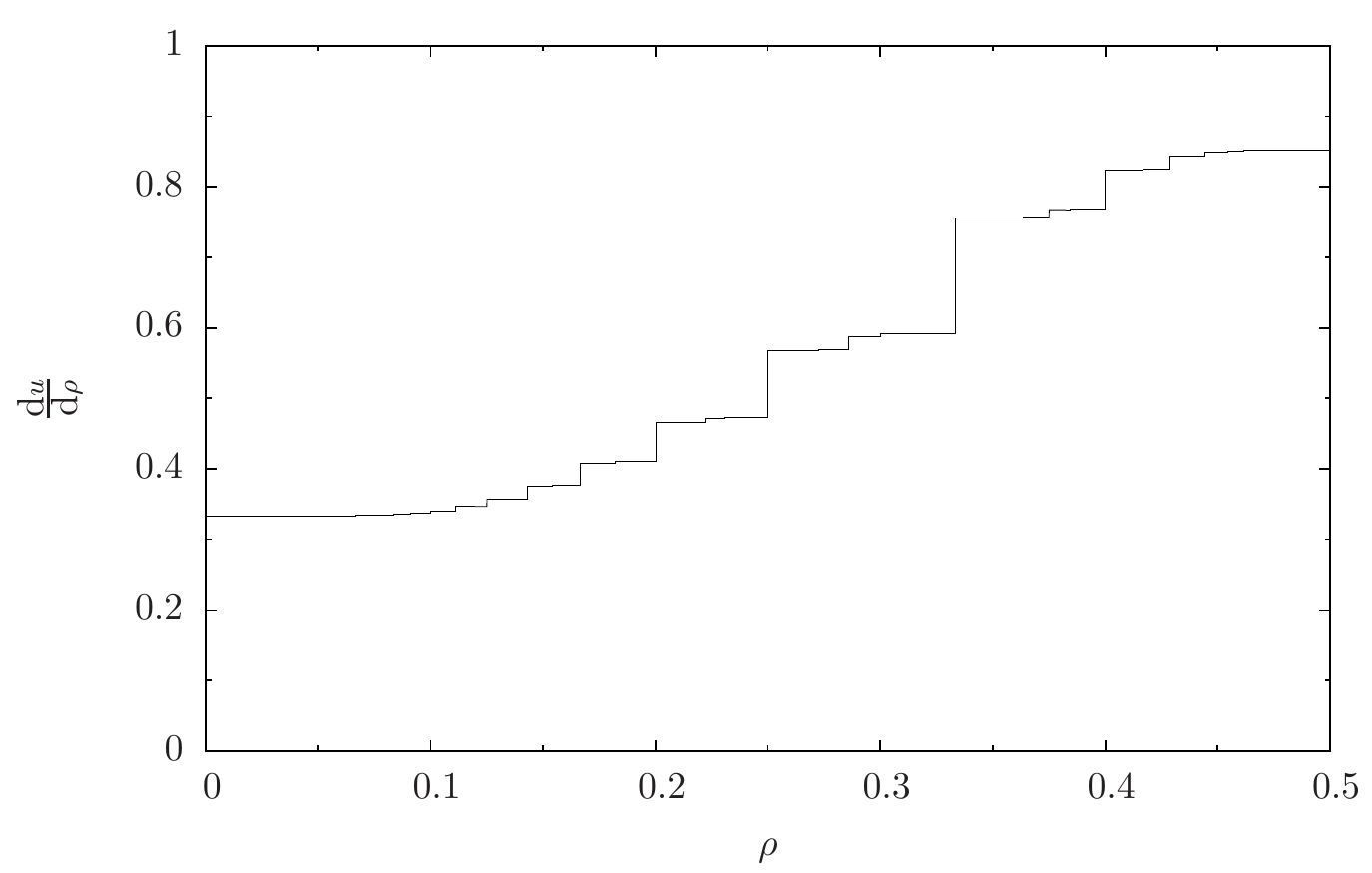}
\end{center}
\caption{\label{fig:wheel}
  Energy $u=h(q_*=0)$ and its first derivative $\text{d}u/\text{d}\rho$
  plotted versus the density $\rho$.
}
\end{figure}

Following Aubry~\cite{aubry1983}, one can obtain an exact expression
for the energy  $U^{min}(\rho)$.   
We need to introduce first another useful representation of the ground
state. If $x_j$ is the position the $j$-th particle, we have
\begin{equation}\label{eq:xj}
x_j = \lfloor \frac {j}{\rho}+ const \rfloor.
\end{equation}
Next, we rewrite the interaction term as
\begin{equation}
  U = \sum_r u_r
\end{equation}
where the $u_r$ is the average contribution of all pairs of $r^{th}$ neighboring particles 
\begin{equation}
u_r = \rho \langle U(x_{j+r} - x_j) \rangle .
\end{equation}
and   $\langle \bullet \rangle$  denotes  averaging over all such pairs, in the limit $M\to\infty$. 
Let us derive an explicit expression for the $u_r$.
The distance $x_{j+r} - x_j$ between two particles separated by $r-1$ others,
is either
$\lfloor r{\rho^{-1}}\rfloor$ or
$\lfloor r{\rho^{-1}}\rfloor + 1$.
If $\pi_r$ is the fraction of pairs  having the value
$\lfloor r{\rho^{-1}}\rfloor$, and $(1-\pi_r)$  of having $\lfloor r{\rho^{-1}}\rfloor + 1$, 
$\langle x_{j+r} - x_j \rangle$ is by definition
$\pi_r \lfloor r{\rho^{-1}}\rfloor
	+ (1-\pi_r)\, (\lfloor r{\rho^{-1}}\rfloor+1)$.
Using the fact that
\begin{equation*}
\langle x_{j+r} - x_j \rangle
  = \langle x_{j+r} - (j+r){\rho^{-1}} \rangle
  - \langle x_{j} - j{\rho^{-1}} \rangle
  + r{\rho^{-1}}
  = r{\rho^{-1}} ,
\end{equation*}
we conclude that 
\begin{equation*}
\pi_r = 1 - \left( r{\rho^{-1}} - \lfloor r{\rho^{-1}}\rfloor\right)
	.
\end{equation*}
We thus obtain an explicit expression of
the interaction term:
\begin{equation}\label{inter}
U^{min}
	= \rho \sum_r   \left[
		\pi_r U(\lfloor r{\rho^{-1}} \rfloor)
		+ (1-\pi_r)\,U(\lfloor r{\rho^{-1}} \rfloor+1)
		\right]
	= \tfrac12  \sum_r
	\left( 2 - \left( r{\rho^{-1}} - \lfloor r{\rho^{-1}}\rfloor\right)\right)\, 
	\rho U(\lfloor r{\rho^{-1}} \rfloor) 
.
\end{equation}
$U^{min}$ is continuous everywhere and its  derivative
\begin{equation}
u'(\rho)= \frac{dU^{min}}{d\rho} 
	= \sum_r \left( 1 + \tfrac12 \lfloor r{\rho^{-1}} \rfloor \right)
		U(\lfloor r{\rho^{-1}} \rfloor)
\end{equation}
is an  increasing function, as shown in Figure ~\ref{fig:wheel}.  

\begin{figure}
\begin{center}
\includegraphics{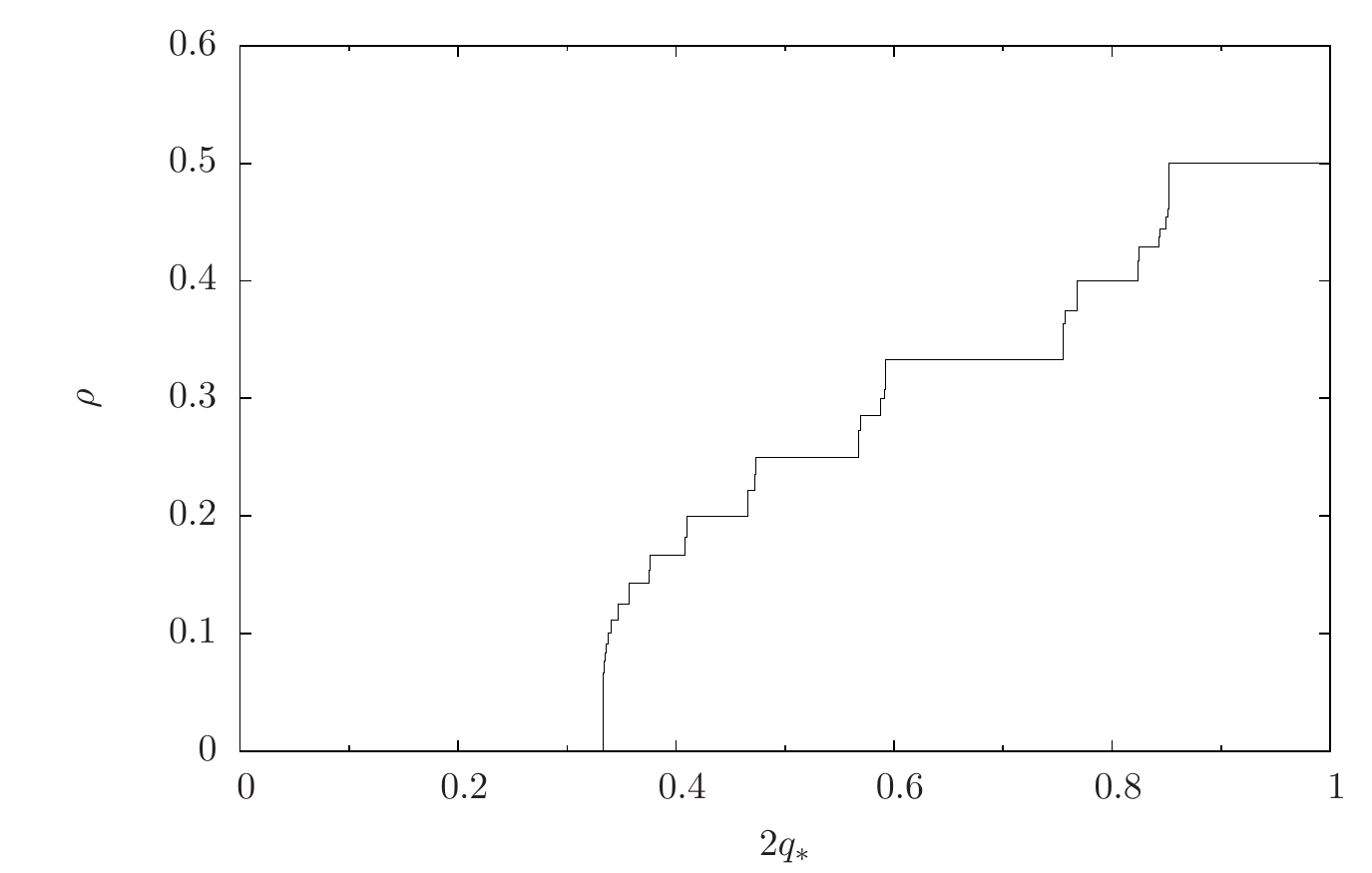}
\end{center}
\caption{\label{fig:devil}
  The density $\rho$ of the ground state plotted versus the chemical
  potential $\mu=2q_*$.  This curve is a complete devil's staircase.
}
\end{figure}

The density can be determined from the knowledge of $q_*$ following equation (\ref{kkk}): the $y$-axis on figure \ref{fig:wheel}
is $2q_*$.  When the level $2q_*$ hits on an interval  between two values, the lower step is chosen. The density $\rho$ is
thus  given by Figure \ref{fig:devil}.
Once $\rho$ is known, the configuration can be obtained either graphically from Figure \ref{fig:bresenham}, or with equation
(\ref{eq:xj}). The value of the 'energy'  can be obtained from  Equations (\ref{eq:h2}) and (\ref{inter}), or the first of figures \ref{fig:wheel}.

If $2q_*$ falls between two steps of the $u'(\rho)$ curve, \textit{i.e.}
there exists a $\rho$ such that $u'(\rho^-) \le \mu \le u'(\rho^+)$, then
$\rho$ is rational and is the density of the ground state; the configuration
of particles is consequently periodic. The width of the interval of $q_*$ corresponding to this density is given by
the discontinuity of the energy's slope $2\delta q_*(\rho) = u'(\rho^+) -
u'(\rho^-)$
If, on the contrary,  there exists a $\rho$ such that $u'(\rho)=2q_*$, then $\rho$ is irrational: the ground state has density
$\rho$ and the configuration of particles is quasi-periodic.

 There exists a smallest integer $r$ such that $r{\rho^{-1}} =
s$ where $s$ is an integer. For this $r$ and for all its multiples $kr$, 
$u'_{kr}(\rho)$ is discontinuous and
$\sum_{k,r} (u_{kr}'(\rho^+)-u_{kr}'(\rho^-) )$
is the sum of the discontinuities. We obtain
\begin{equation*}
\delta\mu\left( \rho=\frac{r}{s} \right)
	= \frac13 \sum_{k=1}^\infty (ks) 2^{-ks}
	= \frac{s}3 \frac{2^{s}}{(2^{s}-1)^2}
	.
\end{equation*}
Note that the result does not depend on the numerator $r$.

Although a quasi-periodic ground state can be obtained by choosing the right
$q_*$, the set of such particular values has zero measure.
Indeed, the sum of all discontinuities of $u'(\rho)$ for $0\le\rho\le 1$
rational, is
\begin{equation*}
\Delta
	= \sum_{\rho=\frac{r}{s}\in[0,1]} \left(
			u'(\rho^+) - u'(\rho^-) \right)
	= \frac13 \sum_{s=1}^\infty \sum_{k=1}^\infty
		\varphi(s) (ks) 2^{-ks}
	,
\end{equation*}
where $\varphi(s)$ is the Euler function equal to the number of
integers in $[1,s-1]$ that are coprime to $s$.
Using the change of variable $u=ks$ and a known property of the Euler
function, we obtain
\begin{equation*}
\Delta
	= \frac13 \sum_{u=1}^\infty
		u 2^{-u} \sum_{k|u} \varphi(k)
	= \frac13 \sum_{u=1}^\infty u^2 2^{-u} = \frac43 .
\end{equation*}
We have proved that $\Delta = \frac43 = u'(1)-u'(0)$, that is, the
probability that a fixed $q_*$ falls between  steps  is one. In other words, the probability that the density is
rational is one and the set of values of $q_*$  associated with
quasi-periodic orbits has zero measure.

The solution $\rho(q_*)$  plotted in figure~\ref{fig:devil} has a devil's staircase structure: it is a \textit{continuous}
curve, constant on some interval when $\rho$ is rational.

\section{Concluding remarks}

We have studied the trajectories of the baker's map that extremize the average squared distance to a line 
 $q=q_*$.
The minimal trajectory is periodic for most values of $q_*$; and quasi-periodic for a zero-measure set of values of $q_*$. 
For all values of $q_*$, the minimal trajectory exhibits more order than the
generic chaotic trajectories. This does not mean that the trajectories involved are not 
chaotic in the sense of being insensitive to boundary conditions: a slight perturbation of the initial condition 
of an extremal trajectory leads to a new trajectory that departs away from the periodic one, and ultimately explores
ergodically all phase-space -- thus spoiling its extremal properties.

The ordered structure of the minimizing state highly depends on fact that the
corresponding Hamiltonian in the 1D lattice gas has infinite range. In the
 case in which the Hamiltonian of the 1D lattice gas has a
finite range ( $U_r=0$ for
$r>R$ --  one may think about a hard-sphere model), for small enough  density any configuration 
with interparticle distance $>R$  minimizes the energy;  and the vast majority of
configurations is not ordered. This is easy to understand for the baker's map: such an interaction corresponds to minimizing a functional that
does not depend on all digits of the positions, i.e. it is a functional in which the trajectories contribute in a degenerate manner.

In the other extreme, functionals that are long-range in time may also yield disordered extremal trajectories.
For the baker's map,  the observable
to be minimized could be
\begin{equation*}
    \lim_{T\to\infty}
    \left(
     \frac1T \sum_{t=0}^{T-1} \left(\lfloor 2q_t \rfloor -1/2\right)
  \right)^2
  ,
\end{equation*}
where $\lfloor 2q_t\rfloor$ is the $t$-th binary digit of $q_0$. This is a long-range antiferromagnet:
the sum  thus has  disordered, degenerate  ground states.  Note that  this observable is not local in the sense that the
position at time $t$ is coupled to the position at different time $t'$.

In conclusion, we have presented a simple example illustrating that those trajectories of a map which extremize some functional may exhibit structure absent in the entire ensemble of trajectories.  This is reminiscent of the Frankel-Kontorova model, where the ground states possess quasiperiodic translational order\cite{Aubry,CF}, while other stationary solutions do not.  It is tempting to conjecture that such behavior is the rule rather than the exception, and may therefore be relevant to systems which are chaotic in either the spatial or temporal domain.

\vspace{8cm}

\pagebreak

\section*{Appendix}

Let us start with 

\begin{equation}\label{eq:Ur}
        U_r^{(N)} = \frac{1}{(2^N-1)^2} \sum_i 2^{i} 2^{i+r} .
\end{equation}
We first determine $U_0^{(N)}$:
\begin{equation}\label{eq:U0}
        U_0 = \frac13 \frac{2^N+1}{2^N-1}.
\end{equation}
Next, let us express 
 $U_{r+1}^{(N)}$ in terms of  $U_r^{(N)}$.
 We split the sums in
\begin{equation}
U_{r+1} = \frac1{(2^N-1)^2} \sum_{i=0}^{N-1} 2^i 2^{[i+r+1]} 
\end{equation}
as
\begin{equation}
\sum_{i=0}^{N-1}
	= \sum_{i+r+1 < N} + \sum_{i+r+1 \ge N}
	= \sum_{i=0}^{N-r-2} + \sum_{i=N-r-1}^{N-1}
.
\end{equation}
so that
\begin{equation}\label{aaa}
U_{r+1}^{(N)} = \frac1{(2^N-1)^2}
	\left(
		\sum_{i=0}^{N-r-2} 2^{i} 2^{i+r+1}
	+	\sum_{i=N-r-1}^{N-1} 2^i 2^{i+r+1-N}
	\right)
\end{equation}
(here the exponents are not necessarily modulo $N$). The first and second sums read:
\begin{equation}
\sum_{i=0}^{N-r-2} 2^{i} 2^{i+r+1}
= 2 \left( \sum_{i=0}^{N-r-1} 2^i 2^{i+r+1} \right) - 2^{N-r-1} 2^{N}
\end{equation}
and
\begin{equation}
\sum_{i=N-r-1}^{N-1} 2^i 2^{i+r+1-N}
= 2 \left( \sum_{i=N-r}^{N-1} 2^i 2^{i+r-N} \right)
+ 2^{N-r-1}
\end{equation}
so that equation  \eqref{aaa} can be written
\begin{eqnarray}
U_{r+1}^{(N)} &=& 2 U_r^{(N)} -
	\frac1{(2^N-1)^2}
	\left(
	2^{N-r-1} 2^{N}
	- 2^{N-r-1} 
	\right)
\\
U_{r+1}^{(N)}	&=& 2 U_r^{(N)} - \frac{2^{N-r-1}}{2^N-1}
\end{eqnarray}
or:
\begin{equation}
\left(2^{r+1}U_{r+1}^{(N)}\right) = 4 \left( 2^r U_r^{(N)}\right)  -  C
\end{equation}
with $C = 1/(1-2^{-N})$.
Putting
\begin{equation}
b_r = 2^r U_r^{(N)} - \frac{C}{3}
\end{equation}
the recurrence equation becomes
\begin{equation}
b_{r+1} = 4 b_r  
\end{equation}
which implies that  $b_r=b_0 4^r$, with $b_0= U_0^{(N)} - \frac{C}{3}$, leading to:
\begin{equation}
U_r^{(N)} = \frac{C}{3} + 4^r \left( U_0^{(N)} - \frac{C}{3}\right)=\frac13 \, \frac{2^{N-r} + 2^r }{2^N - 1} .
\end{equation}

\end{document}